\documentclass[pre,aps,showpacs,twocolumn]{revtex4}
\usepackage{epsfig}
\usepackage{bm}

\begin{document}

\title{Logarithmic scaling in the near-dissipation range of turbulence}

\author{\small K.R. Sreenivasan$^1$, A. Bershadskii$^{1,2}$}
\affiliation{\small {\it $^1$The Abdus Salam International Center
for Theoretical Physics, Strada Costiera
11, 34100 Trieste, Italy}\\
{$^2$ICAR, P.O.\ Box 31155, Jerusalem 91000, Israel}}

\begin{abstract}
A logarithmic scaling for structure functions, in the form $S_p
\sim [\ln (r/\eta)]^{\zeta_p}$, where $\eta$ is the Kolmogorov
dissipation scale and $\zeta_p$ are the scaling exponents, is
suggested for the statistical description of the near-dissipation
range for which classical power-law scaling does not apply. From
experimental data at moderate Reynolds numbers, it is shown that
the logarithmic scaling, deduced from general considerations for
the near-dissipation range, covers almost the entire range of
scales (about two decades) of structure functions, for both
velocity and passive scalar fields. This new scaling requires two
empirical constants, just as the classical scaling does, and can
be considered the basis for extended self-similarity.

\end{abstract}

\pacs{47.27.-i, 47.27.Gs, 47.27.Nz}

\maketitle

The Kolmogorov approach to the phenomenological description of the
inertial range of scales in turbulence requires that the scales $r
\gg \eta$, where $\eta$ is Kolmogorov dissipation scale \cite{my}.
The dominant physical mechanism operating in the inertial range
can be thought to be the Kolmogorov-Richardson cascade, and its
application readily yields the well-known power-law scaling
\cite{frisch,sa}. On the other hand, the near-dissipation range of
scales, for which $r > \eta$ but there exists no separation
between the inertial and dissipation ranges, has a more complex
dynamics arising from a strong competition between the cascade and
dissipation mechanisms (see, for instance, Refs.\
\cite{nelkin},\cite{meneveau},\cite{cgl} and the references cited
there). Indeed, for small and moderate Reynolds numbers, the
near-dissipation range can span most of the available range of
scales. Analogous situation occurs also for the turbulent mixing
of passive scalars.
\begin{figure} \vspace{-0.7cm}\centering
\epsfig{width=.45\textwidth,file=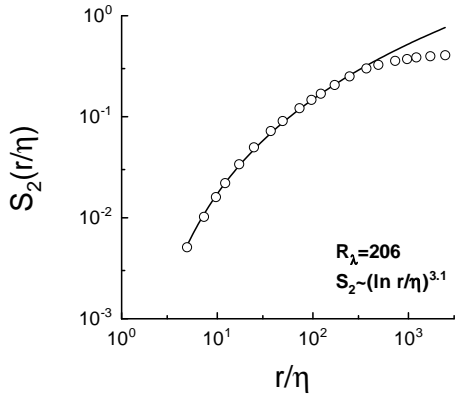} \vspace{-5cm}
\caption{The second order structure function $S_2(r/\eta)$ against
$r/\eta$. The experimental data ($R_{\lambda}=206$) are shown as
circles. The solid curve is the best fit of (3) to the data,
corresponding to the logarithmic scaling. }
\end{figure}
\begin{figure} \vspace{-1.3cm}\centering
\epsfig{width=.45\textwidth,file=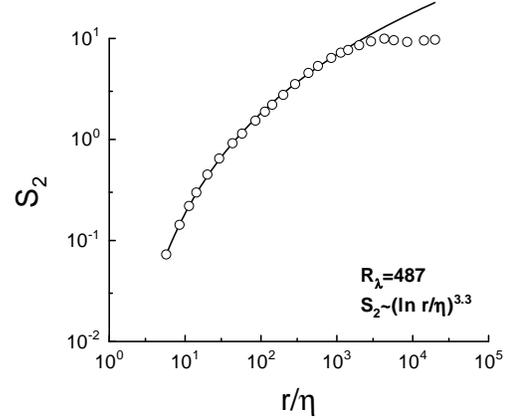} \vspace{-4.5cm}
\caption{The same as in Fig.\ 1 but for $R_{\lambda} = 487$.}
\end{figure}
\begin{figure} \vspace{-0.5cm}\centering
\epsfig{width=.45\textwidth,file=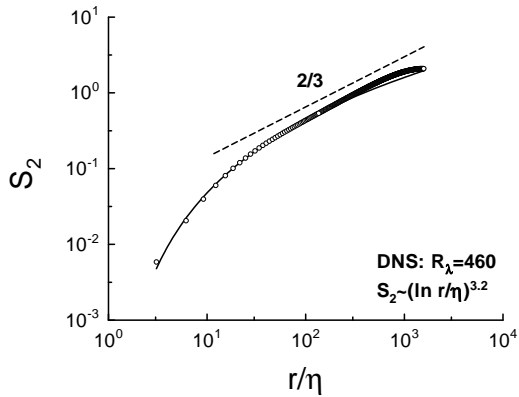} \vspace{-5cm}
\caption{The second order structure function $S_2(r/\eta)$ against
$r/\eta$. The DNS data \cite{gfn} ($R_{\lambda}=460$) are shown as
circles. The solid curve is the best fit of (3) corresponding to
the logarithmic scaling. The dashed straight line is the
Kolmogorov's 2/3-rds scaling form, nominally thought to be
applicable in the inertial range (ignoring effects of small-scale
intermittency). }
\end{figure}

One of the few properties known about the near-dissipation region
is that it obeys the so-called Extended Self-Similarity (ESS)
\cite{ben}, while the classical power-law scaling does not exist.
In ESS the scaling relation between structure functions of
different orders is given by
$$
S_p (r) \sim S_q^{\beta_{p,q}} (r),   \eqno{(1)}
$$
where $\beta_{p,q}$ is the exponent of the $p$-th order structure
function relative to that of the $q$-th order. For the velocity
field in the near-dissipation range, ESS implies that
$$
S_p (r) = f(r/\eta)^{\zeta_p},    \eqno{(2)}
$$
where $f(x)$ is an unknown function different from a power law.
Finding this function is crucial for the near-dissipation range.
We show here, using experimental and numerical data, that the
logarithmic function
$$
S_p \sim [\ln (r/\eta)]^{\zeta_p}   \eqno{(3)}
$$
can successfully replace the power law $S_p \sim r^{\zeta_p}$ in
the near-dissipation range.
\begin{figure} \vspace{+0.3cm}\centering
\epsfig{width=.45\textwidth,file=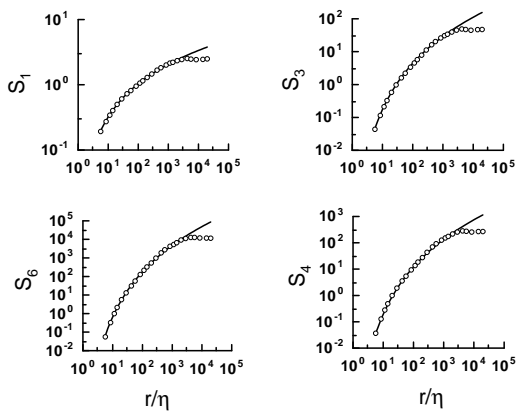} \vspace{-5.3cm}
\caption{Structure functions of different orders ($p=1,3,4$ and 6)
for $R_{\lambda}=487$ (cf.\ Fig.\ 2). The solid curves are drawn
to indicate the logarithmic scaling (3). }
\end{figure}
\begin{figure} \vspace{-1cm}\centering
\epsfig{width=.45\textwidth,file=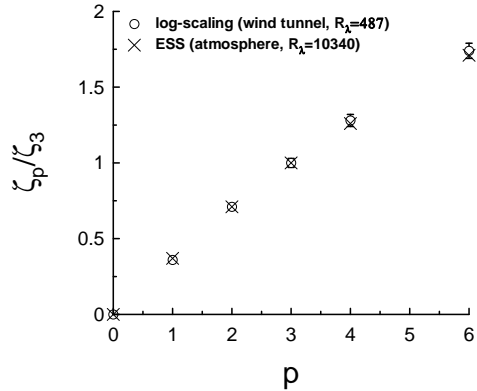} \vspace{-4.5cm}
\caption{Normalized exponents $\zeta_p/\zeta_3$ against p for
$R_{\lambda}=487$ (circles). Crosses are the ESS exponents
obtained for the atmospheric turbulence data at $R_{\lambda}
=10340$ \cite{sd}.}
\end{figure}
\begin{figure} \vspace{-0.2cm}\centering
\epsfig{width=.45\textwidth,file=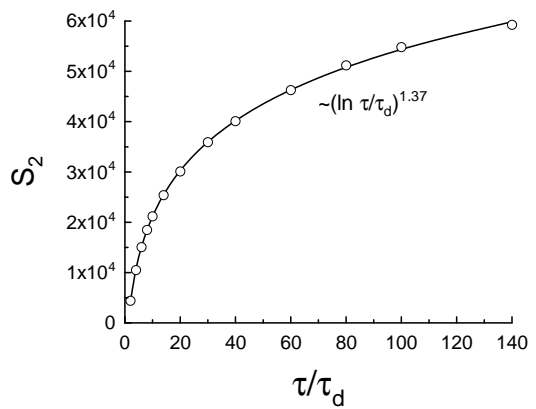} \vspace{-4.5cm}
\caption{Second order structure function of the temperature
fluctuations for the heated wake. }
\end{figure}
\begin{figure} \vspace{-1.2cm}\centering
\epsfig{width=.45\textwidth,file=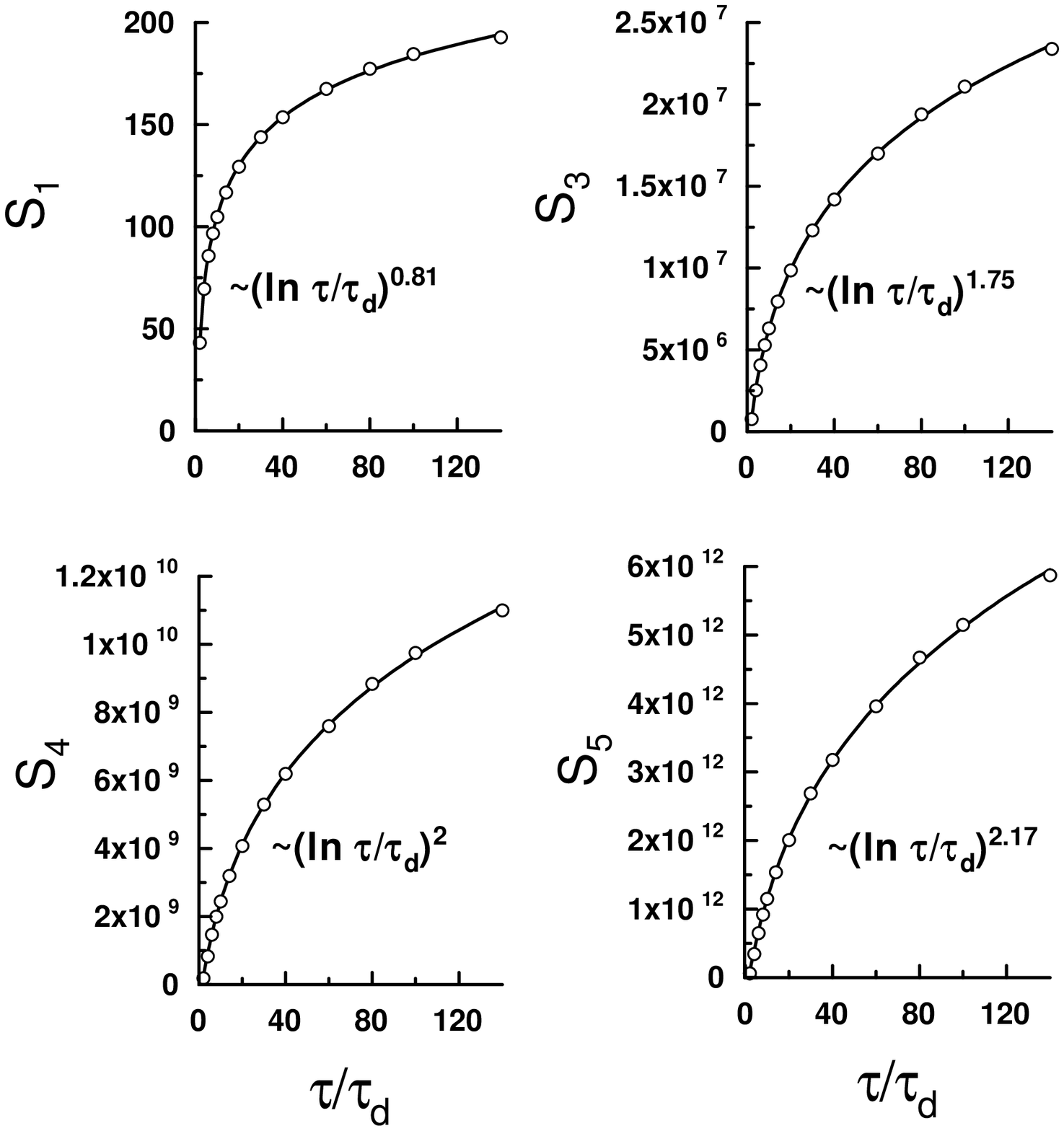} \vspace{-3cm}
\caption{As in Fig.\ 3 but for different orders of the structure
function.}
\end{figure}

Figures 1 and 2 show the longitudinal second order structure
function of the velocity field calculated using the data obtained
in a wind tunnel at $R_{\lambda} = 206$ and 487, respectively.
Here $R_\lambda $ is the so-called Taylor microscale Reynolds
number which varies as the square root of the large-scale Reynolds
number. The flow was a combination of the wake and homogeneous
turbulence behind a grid and is fully described in Ref.\
\cite{pkw}. Following convention, we invoke Taylor's hypothesis
\cite{my} to equate temporal statistics to spatial statistics. We
normalize the scale $r$ by the Kolmogorov scale $\eta$. The solid
curves in these figures are the best fit by equation (3) for
$\zeta_2 \simeq 3.1$ and 3.3, respectively. For comparison, we
show in Fig.\ 3 the quantity $S_2 (r)$ calculated using data from
a high-resolution direct numerical simulation of homogeneous
steady three-dimensional turbulence \cite{gfn}, corresponding to
$1024^3$ grid points and $R_{\lambda}=460$. The solid curve in
this figure also corresponds to the best fit by equation (3), with
$\zeta_2 \simeq 3.2$. The logarithmic scaling applies well to the
near-dissipation region and, although we don't particularly expect
it, to a considerable part of the inertial range as well,
overlapping with the 2/3-rds form of Kolmogorov.

\begin{figure} \vspace{-0.4cm}\centering
\epsfig{width=.45\textwidth,file=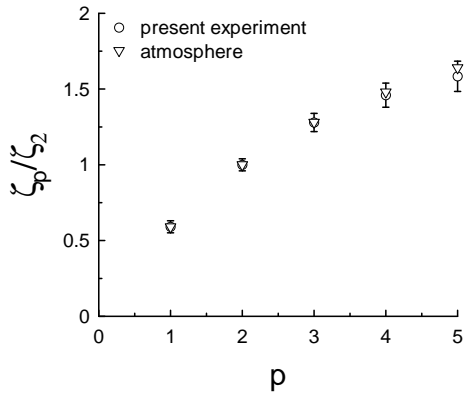} \vspace{-5cm}
\caption{Exponents for the passive scalar extracted from Figs.\ 6
and 7 (circles) and for the fully developed atmospheric turbulence
(ordinary scaling in the inertial interval, triangles
\cite{lov}).}
\end{figure}

Figure 4 shows the structure functions of different orders for
wind-tunnel data ($R_{\lambda} = 487)$ and the logarithmic scaling
(3) is shown in the figure as the solid curves. The exponents
$\zeta_p$ normalized by the exponent $\zeta_3$ are shown in Fig.\
5 as circles. For comparison we show by crosses in this figure the
normalized exponents obtained using ESS in the atmospheric
turbulence for large $R_{\lambda} = 10340$ \cite{sd}. As described
at some length in \cite{sd}, the ESS exponents are very close to
those determined directly.

\begin{figure} \vspace{+0.1cm}\centering
\epsfig{width=.45\textwidth,file=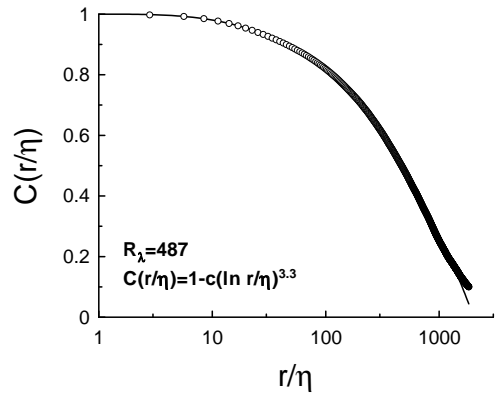} \vspace{-4.5cm}
\caption{Correlation function for velocity fluctuations against
$r/\eta$ for $R_{\lambda} = 487$ (circles). The solid curve is
drawn in the figure to indicate agrement with the logarithmic
scaling shown in Fig.\ 2.}
\end{figure}

The fact that the logarithmic exponents are also the same as the
power law exponents clearly shows why the ESS works well at low
Reynolds numbers. We understand that this observation meerly
shifts emphasis on explaining why these two sets of exponents
should be identical, but leave this question for the future.

For the passive scalar we use the data acquired with a cold-wire
probe in the wake of a heated cylinder in a wind tunnel (see Ref.
\cite{kailasnath} for details of the experiment). Temperature can
be considered a passive scalar for the conditions of the
experiment. We used measurements on the centerline of the wake.
Figures 6 and 7 show the temperature structure functions of
different orders and the solid curves  are the best fits drawn to
indicate agreement with equation (3) (i.e., logarithmic scaling).
Values of $\zeta_p/\zeta_2$ calculated from Figs.\ 6 and 7 are
shown in Fig.\ 8 as circles. We also show by triangles in this
figure the ESS exponents obtained for passive scalar in fully
developed atmospheric turbulence using {\it ordinary} scaling in
the {\it inertial} range of scales \cite{lov}.

It should be pointed out that the logarithmic scaling requires two
fitting constants just as the ordinary scaling does, and, in the
examples discussed above, the range of scales covered by this
scaling is about two decades. Application of these results to
correlation functions and spectra is in good agreement with the
structure functions analysis (see, for instance, Fig.\ 9) as well
as with other available data sets.

Thus, we have clear experimental indication that the logarithmic
scaling is an appropriate tool for the description of data in the
near-dissipation range. This scaling applies to nearly the entire
range of scales in flows of low and modest Reynolds numbers. At
high Reynolds numbers, it overlaps with the conventional power
laws. The logarithmic exponents are the same as the classical
ones, thus demonstrating why ESS works so well at low and moderate
Reynolds numbers. These comments apply to both velocity and scalar
fields. The theory concerning these observations will be discussed
elsewhere \cite{bershads}.

We thank Bruce Pearson and Toshi Gotoh for sharing their data.

\end{document}